\documentclass[preprint,showkeys,preprintnumbers,amsmath,amssymb]{revtex4}

\usepackage{verbatim}
\usepackage{graphicx}% Include figure files
\usepackage{dcolumn}% Align table columns on decimal point
\usepackage{bm}% bold math

\newcommand{\bra}[1]{\langle #1|}
\newcommand{\ket}[1]{|#1\rangle}

\newcommand{\nuoli}[1]{\rightarrow}
\newcommand{\abs}[1]{\left| #1 \right|}

\begin{document}

\title{Characteristics of the polymer transport in ratchet systems}

% PACS: 05.60.Cd, 05.40.Fb, 02.60.Pn, 87.16.Nn
% http://www.aip.org/pacs/pacs2010/individuals/pacs2010_regular_edition/reg80.htm

\author{Janne Kauttonen}
 \affiliation{Department of Physics, University of Jyv\"askyl\"a, P.O. Box 35, FI-40014 Jyv\"askyl\"a, Finland}
 \email{janne.kauttonen@jyu.fi}
\author{Juha Merikoski}
 \affiliation{Department of Physics, University of Jyv\"askyl\"a, P.O. Box 35, FI-40014 Jyv\"askyl\"a, Finland}  
  
\date{\today}

\begin{abstract}

Molecules with complex internal structure in time-dependent periodic potentials are studied by using short Rubinstein-Duke model polymers as an example.  We extend our earlier work on transport in stochastically varying potentials to cover also deterministic potential switching mechanisms, energetic efficiency and non-uniform charge distributions.  We also use currents in the non-equilibrium steady state to identify the dominating mechanisms that lead to polymer transportation and analyze the evolution of the macroscopic state (e.g., total and head-to-head lengths) of the polymers.  Several numerical methods are used to solve the master equations and nonlinear optimization problems.  The dominating transport mechanisms are found via graph optimization methods.  The results show that small changes in the molecule structure and the environment variables can lead to large increases of the drift.  The drift and the coherence can be amplified by using deterministic flashing potentials and customized polymer charge distributions.  Identifying the dominating transport mechanism by graph analysis tools is found to give insight in how the molecule is transported by the ratchet effect.

\end{abstract}


\maketitle

\section{Introduction}

Theoretical research on Brownian motors and the ratchet effect has flourished since the early 1990's \cite{Reimann,Astumian,Wang1,Marchesoni}.  Most studies have been limited to simple systems with one or two coupled particles, whereas research of more complicated systems has escalated in recent years \cite{Downton,Kenward,Gehlen,Mateos,Porto,Eichhorn,Prost,Kazunarib,Ming,Cao,Bao}.  Due to the increased complexity of the models with internal structure, numerical methods play a more important part.  This is due to the fact that the ratchet effect occurs in a far from equilibrium environment and only simple model systems can be analyzed exactly (see, e.g., \cite{Sokolov,Jarzynski}).

The internal structure is an important aspect for many real-life molecular motors (e.g., the well-studied kinesin \cite{Astumian1}).  For systems with non-homogeneous potentials, internal states usually play a more important part than in ``traditional'' transport driven by biased external forces (such as a constant electric field).  Even single particle systems based on the ratchet effect have been shown to display many phenomena, of which the current inversion phenomenon is one of the most interesting.  Current inversions are found to be rather common and can usually be generated by tuning of variables (e.g., diffusion constant, friction, potential shape and/or period) \cite{Reimann,Kostur,Miura1,Miura2,Craig,Chen,Pascual,Grier,Klumpp}.  In view of this it is reasonable to assume that systems with internal dynamics possess even more surprising properties, and knowledge of the correlation between internal states and transport would enable artificial engineering of the molecules and to boost wanted properties such as the velocity. An intriguing possibility considered in this Article is the control of electrophoresis \cite{Electrophoresis} by modifying the internal charge distribution of the molecule. Because of the large number of parameters and different models, it is hard to compare results from different works and form any universal rules about the current or energetic properties for the ratchet effect.  Things get even more complicated for complex molecules, for which the results are even more model dependent.  Therefore we think that it is necessary to at least develop some general methodology for how to systematically study and monitor the behavior of these systems.  This is indeed one of the key themes of this work.

We have recently studied polymers using the Rubinstein-Duke (RD) model in time-dependent periodic potentials \cite{Kauttonen}.  The RD model  \cite{RubinsteinDuke} is a good prototype of a complex molecule since the size of a linear polymer can be easily varied, it is strongly correlated, and the model has been actively studied for two decades \cite{Defontaines,Leeuwen,Kooiman,Sartoni,Carlon,Kolomeisky1,Constrait}.  There has also been interest towards polymers as Brownian motors recently \cite{Kulakowski,Downton,Kenward}.  In Ref.~\cite{Kauttonen} we presented a general ``toolbox'' based on the numerical solutions for master equations and found current inversions for the RD model in the flashing ratchet and traveling potentials.  In this Article we extend our work and methodology by considering the efficiency, different potential time-dependency schemes, non-homogeneous charge distributions and the dynamics of the internal states leading to the macroscopic transport properties.  We formulate the operators and master equations that are then solved with suitable numerical tools that fall into areas of linear algebra, integration, optimization and graph analysis.  Due to the nature of the ratchet effect, most observables that we are interested in (such as drifts and conformational changes) are very small.  Therefore we find that a discrete space model that allows numerically exact solutions provides a very useful framework in this context.

It is found that, like in many other studies on the ratchet effect before, varying certain model parameters has a large effect on the velocity, coherence and energetic efficiency.  We take this aspect a step further by doing multiple parameter optimization for the RD model in order to maximize the steady state drift.  If the internal states and the movement of the polymer are tightly correlated (such as in the RD model), changing the parameters increases the importance of some molecule conformations over the others.  We demonstrate this by comparing the expected values for certain characteristic macroscopic properties for polymers, such as head-to-head and total length.  We also identify and compare the most important microscopic conformations of the polymer that are responsible for the currents in different situations.

This Article is organized as follows.  In Section II a mathematical framework and notations are defined and in Section III we go through the numerical methodology.  In Sections IV and V we present our results for transport properties and examine their microscopic origin.  In the Appendix, the operator algebra involved is discussed in detail.  Our conclusions are given in Section VI.

\section{Model}

We study the transport of the RD polymer \cite{RubinsteinDuke} and its modification, the \emph{free motion} (FM) polymer \cite{Kauttonen}, in temporally and spatially changing driving potentials.  Essentially the RD model consists of connected Markovian random walkers (\emph{reptons}) in continuous time (see Fig.~\ref{fig:esimerkkikuva}).  Each repton carries a charge that interacts with the potential.  The model was originally developed to study the reptation process of the polymer in a restrictive medium (gel).  However, in the context of this study, the model is primarily used as a good prototype of a molecule with a large number of internal states.  To study the importance of the bulk motion, the assumption of the reptation can be relaxed, which results in the FM model.  The complexity of the polymers can be increased by considering arbitrary charges of the reptons.  In the following Section, formal definitions of the model are given for the implementation for numerical computations.  Readers not interested in the technical details may skip this part and proceed to Section B.

\subsection{Stochastic generator and operators}

Consider a one-dimensional discrete Markovian random process in continuous time \cite{Kampen}.  After the transition rates between all the allowed states ($i,j$) in the system are given (elements $H_{i,j}$), the stochastic matrix $H$ can be defined.  For molecular motors, this matrix includes all the internal conformations and spatial positions of the molecule in the potential \cite{Fisher,Maritan,Kolomeisky}.  In the case of the Markovian stochastically driven potential, it also includes the states of the external potential.  We consider systems with stochastic (\emph{type 1}) and deterministic potential switching schemes with sudden (\emph{type 2}) and smooth (\emph{type 3}) switching.  The potential $V(x,t)$ is assumed to be $L$ and $T$ periodic in space and time (for stochastic switching $T$ is the expectation value).  The stochastic matrix for the polymer dynamics is
\begin{equation*}
% H(\mathbf{q},E,T) =
H = \sum_{s=1}^{S} \left[ \sum_{l=1}^{L} \left[ A_{l,s} + \sum_{y} \left( B_{y,l,s} + \sum_{i=1}^{N-2} M_{i,y,l,s} \right) \right] + \frac{1}{T_{s}} \left( \widehat{n}_s - \widehat{h}_{s} \right) \right],
\end{equation*}
for the type 1 and
\begin{equation}
H(t) = \sum_{l=1}^{L} \left[ A_{l}(t) + \sum_{y} \left( B_{y,l}(t)+\sum_{i=1}^{N-2} M_{i,y,l}(t) \right) \right] ,
\end{equation}
for the deterministic case, where $T_{s}$ is the expected lifetime of the potential $V_s$, and $\mathbf{q} \in \Re ^N$ the repton charges.  The switching of the potential is assumed to be cyclic, i.e., $V_1 \rightarrow V_2 \rightarrow \dots \rightarrow V_s \rightarrow V_1$.  The operators $\widehat{n}_s$ and $\widehat{h}_{s}$ create transitions between the potential states, and the operators $A$, $B$ and $M$ determine the dynamics of the head ($A$ and $B$) and middle ($M$) reptons.  The detailed definitions of these operators are given in the Appendix (see also Ref.~\cite{Kauttonen}).  In Fig.~\ref{fig:esimerkkikuva} we have an illustration of the six repton polymer in one of its configurations.  We fix the direction of the motion such that up arrows indicate the positive direction and vice versa.

\begin{figure}
\includegraphics[width=10.0cm]{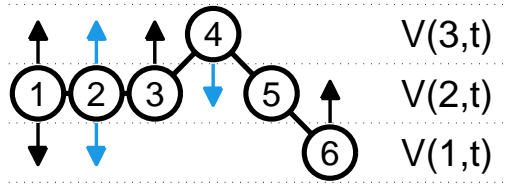}
\caption{(Color online) Illustration of a single configuration of the six repton ($N=6$) polymer in the external potential $V(x,t)$ 
with available moves shown by arrows.  The moves with blue (gray) arrows are only available in the FM model.}
\label{fig:esimerkkikuva}
\end{figure}

The type 2 operator now becomes
\begin{equation*}
H(t)=
\begin{cases}
H_{\rm 1} \: , \quad t \in \left[ 0,T_1 \right)
\\ H_{\rm 2} \: , \quad t \in \left[ T_1,T_1+T_2 \right)
\\ \vdots
\\ H_{\rm S} \: , \quad t \in \left[ \sum_{i=1}^{S-1} T_i,T \right) ,
\end{cases}
\end{equation*}
and for the type 3 we choose $V(x,t)=V_{\rm max}(x) \sin^2 (\pi t/T)$.  The type 2 and 3 potentials are more reasonable for artificial molecular motors that have external driving mechanisms (e.g., electric potential), whereas the type 1 occurs most likely in nature (e.g., ATP driven motors).  After the generator is defined, the dynamics is given by the master equation $d P(t)/dt = H(t) P(t)$, where the elements of the probability vector $P(t)$ include all the individual states $y$ of the system.  The stationary state $P_{\rm stat}$ for the type 1 generator means that $H P_{\rm stat}=0$ and for types 2 and 3 that $P_{\rm stat}(t)=P_{\rm stat}(t+T)$.  From $P_{\rm stat}$, all expected values, such as the drift $v$, can be computed.  The effective diffusion coefficient $D_{\rm eff}$ is found by solving another stationary state for the diffusion equation (see Ref.~\cite{Kauttonen}).  After solving $v$ and $D_{\rm eff}$, the Peclet number can be computed from $\text{Pe} =  \frac{ \abs{ v }}{D_{\rm eff}}$.  

Although the drift and Peclet number are the most studied properties, they tell nothing about the internal dynamics of the molecule.  More specific measurements are needed.  For a discrete model, individual states, which we call \emph{microstates}, can be bunched together to define the \emph{macrostates}.  Within the operator formalism, the general form of such macrostate operator is
\begin{equation}
\widehat{O} = \sum_{k} \omega_k \sum_{y \in F_k} \widehat{n}_y \\
\label{eq:general_op}
\end{equation}
where $\omega_k$ is the corresponding value of the macrostate (e.g., the polymer length), $\widehat{n}_y$ a microstate operator, and $F_k$ a (large) collection of microstates.  For the RD-type model, there are $3^{N-1}$ microstates, for which the operators have the form
\begin{equation*}
\widehat{n}_y = \prod_{i=1}^{N-1} n_{g(y,i)},
\end{equation*}
where the function $g(y,i)$ defines the state ($A$, $B$ or $\varnothing$) of each bond $i$ between the reptons $i$ and $i+1$.  We define the following four macrostate operators for the RD-type model: the \emph{zero-bond count} (number of $\varnothing$-bonds), the \emph{kink count} (number of $AB$ or $BA$ bond pairs), the \emph{head-to-head length} (distance between first and last repton) and the \emph{total length} (maximum distance between two reptons).  The head-to-head and total lengths are calculated in the potential direction (the only spatial direction for the one-dimensional model) and for the fully accumulated polymer they both are zero.  The corresponding operator definitions of these observables are found in the Appendix.  Separating the head-to-head and the total length is important since the polymer can take a U-shape.  For example, for the configuration in Fig.~\ref{fig:esimerkkikuva} the values for $\omega_k$ of these operators would be 2 for zero-bonds (formed by reptons 1-3), 1 for kinks (reptons 3-5), 1 for the head-to-head length and 2 for the total length.

\subsection{Selection of the rates}

\begin{figure}
\includegraphics[width=10.0cm]{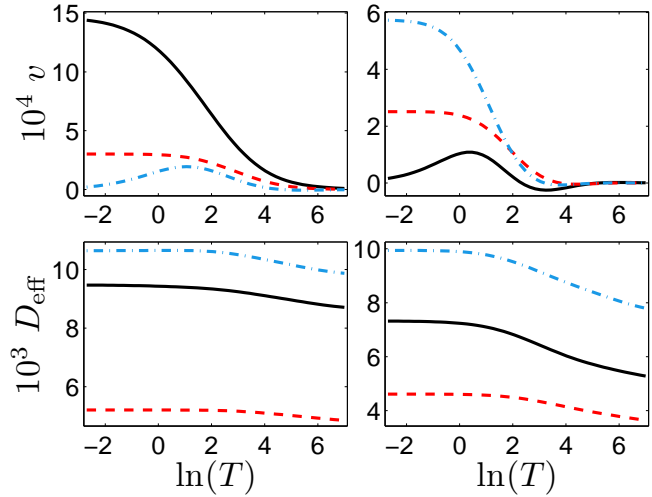}
\caption{(Color online) The effect of the jump rate scheme.  The drift velocity and the diffusion coefficient of the 8-repton RD polymer as a function of the temporal period $T$ in the type 1 flashing ratchet (left) and the traveling potential (right), with exponential (blue dash-dotted lines), Metropolis (black solid lines), and Kawasaki (red dashed lines) rates.}
\label{fig:ratevertailu}
\end{figure}

Despite the large number of studies with discrete state Brownian motors, the importance of choosing the rates $H_{i,j}$ has not got much attention.  By demanding the local detailed balance (no net currents in equilibrium), the usual choices for the rates are \cite{Kramers,Kawasaki}
\begin{equation*}
\frac{1}{\Gamma}H_{i,j}= 
\begin{cases}
 \begin{array}{ll}
	\min \left\{ 1, e^{\left( E_j-E_i \right) / k_B T } \right\} & \text{(Metropolis)} \\
	e^{\left( E_j-E_i \right) /2 k_B} & \text{(exponential)} \\
	\left[ 1 + e^{\left( E_i-E_j \right) / k_B T} \right]^{-1} & \text{(Kawasaki)}
	\end{array}
\end{cases}
\end{equation*}
where $\Gamma$ sets the time-scale and $1 / k_B T$ is the Boltzmann factor.  Both of these constants and the lattice constant, are set to $1$ in this paper.  All three definitions lead to the required $P_i=\exp(-E_i)/Z$ distribution in equilibrium, but generate the different kinds of dynamics when applied to ratchet systems (far from equilibrium) such that the microstate energy $E_i$ contains the potential.  To demonstrate this, we have plotted in Fig.~\ref{fig:ratevertailu} the stationary state drift and diffusion coefficient of the 8-repton RD polymer in the type 1 flashing ratchet and traveling potential (model parameters are listed in the beginning of the section IV).  Although all three curves for flashing and traveling potentials share a similar shape, the scales are different and large differences can be seen in the limit where the temporal period $T \rightarrow 0$.  Being fast and simple, the Metropolis form is usually the favorite choice for the rates.  But especially with ratchet systems it can be a poor choice since it does not take into account the slope of the downhill moves (rate being limited to 1) that is important for the dynamics.  This is also true for the Kawasaki form, since it is basically just a smoothened Metropolis function.  Since there is no single correct choice for the rates (based on theory), the selection must be made on experimental or model specific grounds.  Only exponential (in flashing ratchet) and Metropolis (in traveling potential) dynamics lead to zero drift in this limit, which is a physically more realistic situation and is also consistent with the single Brownian particle model \cite{Bier}.  Therefore we choose these rates in this study.

\subsection{Non-uniform charge distributions}

The usual assumption in the studies concerning polymer transport is that all monomers are identical, i.e., they carry identical charge and mass.  We relax this assumption and study the effect of the non-uniform charge distributions along reptons.  Previous works on the RD model have considered some aspects of this.  In Ref.~\cite{Carlon}, a magnetophoresis model (i.e., one charged head repton) was considered and in Ref.~\cite{Leeuwen1} it was shown that when it comes to the drift velocity all charge distributions are equivalent in small fields (i.e., linear response regime) \footnote{This is clear, since in the linear response regime only the total force directed to the object counts and the diffusion coefficient is given by the Nernst-Einstein relation.  Similar equivalence does not hold for systems involving non-homogeneous potentials.}.  In Ref.~\cite{Nonhomogeneous} it was noticed that the drift in constant field depends strongly on the position of the charged repton within the polymer and in Ref.~\cite{Kolomeisky1} non-homogeneously charged RD polymers in large fields were studied.  Recent study of the dimer in the periodic potential show that if the connected particles are non-identical, directed drift can be generated even in the symmetric potential \cite{Gehlen}.

We want to find the best possible charge distributions $\mathbf{q}$ for the RD and FM polymers by finding the largest possible drifts.  This leads to a multi-dimensional, nonlinear constrained optimization problem with constrains coming from the charges $q_i$.  We choose $\sum_i q_i=Q$ and $q_i \geq 0$, where the first constraint simply sets the total charge corresponding to an uniformly charged polymer and the second one fixes the sign of the charges.  The optimal charge distribution gives some (indirect) information about the polymer conformations and reptons that dominate the transport (i.e., have the largest impact on the drift).  Lastly we note that optimization has been carried out for some single particle systems \cite{Berger,Dinis,Zhou}.

\section{Numerical methods}

\subsection{Network analysis}

The stochastic matrix $H$ can be also treated as a graph with vertices (states) and edges (transitions) that can be analyzed to gain more detailed information of the transport process, as described in this section.  Graphs and statistical physics have a long history due to the close similarities between stochastic systems and electric circuits, and in the seminal work of Schnakenberg in 70's many important results between these two were presented \cite{Schnakenberg} (see \cite{Schmittmann} for some recent developments).  Most of the works on this subject deal with the relations between steady state, rates, probability fluxes and entropy.  We are however interested in finding the optimal paths within a current graph, which has not gained interest within previous works.  Such ideas have however risen in other disciplines such as microbiology \cite{Kamp}.  In the following, only basic knowledge of the graph theory is expected (see, e.g., \cite{Graafiteoria}).  For simplicity, we consider only type 1 scheme where the time-dependency of the stationary state does not need to be explicitly dealt with thus making the numerical computations easier.

After the stationary state $P_{\rm stat}$ of $H$ is found, the net currents (edge weights) between the states can be computed.  In addition to the stationary state and stochastic generator, we make use of the matrix $H^{\rm sign}$ which carries the information about the direction and magnitude for the transitions in $H$.  For the RD-type model, the elements of $H^{\rm sign}$ are $\pm a/N$ for all polymer state transitions, where $a$ is the lattice constant and the factor $1/N$ results from the center-of-mass motion, and zero for the potential state transitions (present only for the type 1 system).  For additional details, see Ref.~\cite{Kauttonen}.  The graphs $G$ and $G^{\rm sign}$ are then formed as follows.  Let $w_{i,j} = P_{\rm stat}(i) H_{i,j} - P_{\rm stat}(j) H_{j,i}$ $\forall i,j$.  If $w_{i,j} \geq 0$, there exists a directed edge $i \rightarrow j$ in $G$ and $G^{\rm sign}$ with weights $w_{i,j}$ and $w_{i,j}^{\rm sign}= \pm H_{i,j}^{\rm sign} w_{i,j}$.  With the sign in front of the weights $w_{i,j}^{\rm sign}$, one chooses the direction of interest of the transport (see below).  The weights $w_{i,j}$ are probability flows and the weights $w_{i,j}^{\rm sign}$ are mean displacement flows in the stationary state.

Let $\gamma_k$ be a path $i_1 \rightarrow i_2 \rightarrow \dots \rightarrow i_k$ in the graph with $i_x \neq i_y \forall x \neq y$ (the path is non-intersecting) and $\gamma_k (j)=i_j$.  We then look for the path(s)
\begin{equation}
\max_{\gamma_k} \sum_{i=1}^{k-1} \frac{1}{k} X_{\gamma_k (i+1),\gamma_k (i)} =: \max_{\gamma_k} f(\gamma_k),
\label{eq:network_optimization}
\end{equation}
with $X$ being $w_{i,j}$ or $w_{i,j}^{\rm sign}$.  The resulting path computed with $w_{i,j}$ contains transitions that lead to the largest mean probability flow and we denote it by $\widetilde{\gamma}_k$.  Similarly with $w_{i,j}^{\rm sign}$, one gets the path with the largest mean probability flow and we denote it by $\widetilde{\gamma}^{\rm sign}_k$.  We call these paths the \emph{dominating processes}.  The function $f$ is known as the target function.  If the system is closed (periodic), the process must eventually return to its starting state and a cycle is formed, in which case  $\gamma_k=\gamma_1$.  Since the potentials we study are indeed periodic, we concentrate on closed systems from now on.  For the cycle $\widetilde{\gamma}^{\rm sign}_k$, the target function defines the mean cycle velocity, i.e., $v_{c} = \sum_{i=1}^{k-1} w_{y_{i+1},y_i}^{\rm sign}$, where $y_i = \widetilde{\gamma}^{\rm sign}_k (i)$.

Whether there is a difference between $\widetilde{\gamma}$ and $\widetilde{\gamma}^{\rm sign}$ depends on the details of the system.  It may turn out that $\widetilde{\gamma}$ only includes transitions that are not responsible for the directed molecule transportation, but instead results from the non-transporting diffusive motion.  Formally this means that $\sum_{i=1}^{k-1} H_{\widetilde{\gamma}_k (i+1),\widetilde{\gamma}_k (i)}^{\rm sign}=0$, which we call a \emph{stationary process}, as the net transport for the cycle is zero.  This is indeed typical for the ratchet transport, since the molecule spends most of its time near the minima of the potential, being unable to move until the suitable state of the potential and molecule conformation is reached.  Therefore $\widetilde{\gamma}^{\rm sign}$ carries more interesting information as it takes into account the directions and magnitude of the moves.  If the path has a property $\sum_{i=1}^{k-1} H_{\gamma_k (i+1),\gamma_k (i)}^{\rm sign} \neq 0$, we call it a \emph{transporting process}.  It is not guaranteed that the dominating process is a transporting process in either case.

In the literature, the problem in Eq.~\eqref{eq:network_optimization} for cycles is known as the \emph{optimum cycle ratio problem} (see, e.g., \cite{Dasdan}).  The graphs $G$ and $G^{\rm sign}$ may include all states of the system or a fraction of them with the rest summed over, hence the level of the coarse graining can be chosen.  For example, if one is interested only on the molecule internal dynamics, summing over all states of the potential may turn out useful.  For a RD-type model this would mean that the dimension of the graph is reduced by a factor of $1/SL$, which also makes the numerical optimization easier.

The dominating processes simply give a collection of the most probable transitions that the molecule can go through successively, thus giving information about the types of processes that are important.  The probability for the (complex) molecule to precisely follow such fixed paths is of course very small.  Because of this, it would be hard and time consuming to try to identify dominating processes from the simulation or experimental data.  Our proposed graph analysis is simple and can in principle be done for all finite discrete stochastic non-equilibrium systems which have non-zero currents.  Whether this analysis is worth the effort (i.e., if $\widetilde{\gamma}$ does contain interesting information), depends on the complexity of the system and the importance of the molecule internal dynamics to the transport process.

\subsection{Motor efficiency}

The efficiency of the molecular motor is an important aspect, especially for non-artificial molecular motors that have limited energy available.  In the literature, there are several definitions of the efficiency for Brownian motors, see, e.g., Refs.~\cite{Reimann,Wang2,Cisneros,Derenyi,Oster,Munakata}.  Here we adopt the basic thermodynamic definition that relies on the constant load force $F$ on the polymer, which means that the output power of the motor is $vF$.  The input power $W_{\rm in}$ comes from turning the potential on, thus forcing the polymer periodically in a higher energy state depending on its location.  This approach is different from the model where the molecule gains constant amount of energy by, e.g., ATP hydrolysis.  We assume that the energy is dissipated, when the polymer goes back to lower energy state, i.e., this energy is not taken into account by reducing it from the input energy.  By assuming that transitions between potentials of type 1 system are cyclic (i.e., $V_1 \rightarrow V_2 \rightarrow \dots V_S \rightarrow V_1$), the input power for stochastic and deterministic potential schemes can be written as
%
% esim. parrondon (-98) juttu
\begin{equation*}
W_{\rm in} = 
\begin{cases}
\begin{array}{ll}
	\sum_{s=1}^{S} \sum_{\epsilon_s} \max \left[ 0 , E_{s+1}(\epsilon) - E_{s}(\epsilon) \right] T_s^{-1} P(\epsilon_s) & \text{, type 1} \\
	\sum_{\epsilon} \frac{1}{T} \int_{t=0}^{T} dt \max \left[ 0 , \frac{d E(t,\epsilon)}{dt} \right] P(\epsilon,t)  & \text{, types 2-3}
\end{array}
\end{cases}
\end{equation*}
where $\epsilon$ and every $\epsilon_s$ include $L 3^{N-1}$ states.  Since the type 2 potential has discontinuities in $t$, one can define $dE(t,\epsilon)/dt := \sum_{s=1}^{S} \left[ E_{s+1}(\epsilon) - E_{s}(\epsilon) \right] \delta (t - \sum_{k=1}^{s} T_k)$.  The efficiency is defined by $\eta = \frac{vF}{W_{\rm in}}$.

Although the efficiency of the flashing ratchet model is very low for single particles (see, e.g., Ref.~\cite{Cisneros}), it can be greatly increased for some many particle systems as shown in the recent work \cite{Kazunarib,Ming}.  Besides the efficiency, we are also interested in the stopping force $F_{\rm stop}$ which, when applied, causes the average drift go to zero.  It is expected that the stopping force gets larger as $N$ increases, as seen in Ref.~\cite{Downton}.

\subsection{Algorithms}

When dealing with large linear systems (of the order of $10^{5}$ states and beyond), one must really pay attention to the convergence properties and therefore the choice of the numerical methods are important.  In this Paper we have three types of numerical problems to solve $P_{\rm stat}$.  For the fully stochastic system (type 1) we used the Arnoldi and bi-gradient stabilized (BiGradStab) methods (drift and diffusion), for on/off deterministic system (type 2) adaptive Runge-Kutta 4-5 method and for smooth continuously deterministic system (type 3) quasi-minimal residual (QMR) method.  The solution of the type 1 problem is a straightforward eigenstate computation, the other two are more involved integration problems.  All computations were performed in Matlab with a modern desktop computer. Solving stationary states for the type 2 and 3 potentials were the most time consuming parts of the computations.

When solving the stationary state for type 1, a random initial vector is good enough choice, but for types 2 and 3 this is not the case.  A better initial guess is needed to reduce the computation time.  We found that the stationary state of the mean-field operator ($H_{\rm MF} = \sum_{k} x_k H_{k} \text{ with } x_k=T_k/T$) is easy to compute and a good one to begin with.  In many cases, previous solutions can be also used (e.g., when varying $T$).  A random initial state however serves as a good check of the numerics, since the results must not depend on the choice of the initial state.

The stationary solution for the type 3 can be found with the same manner as for the type 2 (RK45), which however requires that the operator is available for all $t \in \left[ 0,T \right]$ and are either re-build every step or loaded from the memory.  The other way (which we used) is to solve the larger linear equation problem as a first order discretization in time,
\begin{equation}
H(t) \mathbf{P}(t) \approx \frac{\mathbf{P}(t+\Delta t ) - \mathbf{P}(t-\Delta t )}{2 \Delta t},
\end{equation}
where $\Delta t = T/M$, $M$ being the number of discretization steps.  We found that $M=30...60$ is accurate enough.  In the matrix form this leads to the problem $\widetilde{H} \widetilde{\mathbf{P}} = \mathbf{A}$, where $\widetilde{H}$ includes $H(t)$ for all $M$ time-steps and the discretization operator, and the normalization is preserved with $A_i=1 \forall i = LY,2LY,...,MLY$ otherwise zero.  As before, the time-dependent diffusion coefficient is found by solving another linear problem.  For these linear systems the QMR method turned out to be well converging (LSQR is also a fool-proof method, but very slow).  

To maximize or minimize the velocity $v(\mathbf{q},T)$ for charges and the temporal period, nonlinear optimization can be carried out with the standard sequential quadratic programming method.  To find the velocity, the generator $H(\mathbf{q},T)$ must be constructed several hundred/thousand times because of changes in the transition rates.  Efficient implementation presumes that this process is fast, which is achieved for example by manipulating the required matrix elements directly in the memory instead of re-building the whole matrix.  The choice of the initial state is crucial (as usual for optimization problems) and a random state is used with several repetitions to confirm the global optimal point.  A symmetric initial charge distribution easily leads to a local optimal point with a symmetric charge distribution (as seen in Section IV C).  If $\mathbf{q}$ is fixed, optimization can be replaced by interpolation, since function $v(T)$ is very smooth.

The best known exact algorithms to find the optimal cycle ratio have the complexity $O(n m)$ \cite{Karp}, where $n$ and $m$ are number of vertices and edges, but in practice these algorithms are not the fastest ones \cite{Dasdan}.  We applied an improved version of the Howard's method \cite{Howard} implemented in the \emph{Boost} C++ library.  There also exist brute-force methods to efficiently find (enumerate) all cycles in graphs \cite{Tarjan}, but this approach is limited to very small networks and/or cycle lengths.  We also tested a simple greedy algorithm where we begin from a single edge with the largest weight and start to grow the path by always choosing the edge with the largest weight available at the moment, until the path form a cycle (i.e., crosses itself).  This method however works poorly and an optimal solution is found only for very simple cases (e.g., a polymer in strong static field), where the results are also easy to guess beforehand.  In general situations, the optimal path contains transitions that cannot be chosen by a simple greedy algorithm.

\section{Results for the different potential and polymer types}

Since both RD and FM models include a large number of parameters, some of them must be fixed, primarily those that have a minimal qualitative impact on the results.  In addition to $N$ (reptons), other parameters in the models have the following interpretations:

\begin{itemize}
	\item The environment $\leftrightarrow$ the potential $V(x,t)=V(x+L,t+T)$
	\item The medium $\leftrightarrow$ tube deformation $\Omega$ ($0$ for RD, $1$ for FM)
	\item The polymer internal fine-structure $\leftrightarrow$ charges in $\mathbf{q}$
\end{itemize}

The single most important parameter is the period $T$ of the potential, which is also one of the easiest one to control in experimental set-ups.  The parameter $\Omega$ models the porosity and viscosity of the medium by either restricting polymer strictly into the reptation tube ($\Omega = 0$) or not ($\Omega = 1$).  As before in Ref.~\cite{Kauttonen}, we set $S=2$ and $L=3$ to achieve a both maximal $N/L$ ratio and keep feasible matrix sizes.  The flashing ratchet is $V_1(1)=V_{\rm max},V_1(2)=V_{\rm max}/2,V_1(3)=0$ and $V_2(x)=0 \ \forall x$, and the traveling potential $V_1(1)=V_2(2)=V_{\rm max}$ and zero for $V_1(2),V_1(3),V_2(1)$ and $V_2(3)$.  In Fig.~2 of Ref.~\cite{Kauttonen} there is an illustration of these potentials.  Time symmetry parameter $x=T_{1}/T$ is fixed to $1/2$ for the flashing ratchet potential and $1/4$ for the traveling one.  The maximum potential strength $V_{\rm max}$ has only a small effect on the results and is set to unity (with one exception in Fig.~\ref{fig:optimal_phases}) \footnote{Note that the choice $V_{\rm max}=1$ in this paper is equivalent with $V_{\rm max}=1/2$ of the previous paper, where the factor $1/2$ was dropped in the definition of rates $H_{i,j}$}.  The direction of the potentials is set up in such way that the expected ``main drift'' is always positive and the inverse drift (if present) is negative.

With the definitions in Section II, we study the following three types of time dependent potentials

\begin{itemize}
	\item Type 1: stochastic on/off switching
	\item Type 2: deterministic on/off switching
	\item Type 3: deterministic smooth cosine-type modulation.
\end{itemize}

\subsection{Comparison of time-dependency schemes}

\begin{figure}
\includegraphics[width=13cm]{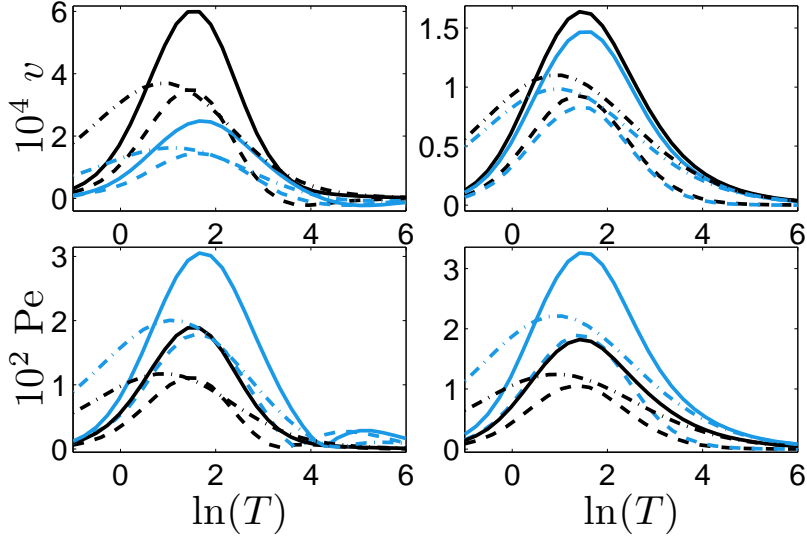}
\caption{(Color online) Drift velocity and Peclet number for $5$ (blue (gray) lines) and $9$-repton (black lines) RD (left) and FM polymers (right), for type 2 (solid lines), type 3 (dashed lines), and type 1 (dash-dotted lines) schemes.}
\label{fig:schemes}
\end{figure}

First we compare the differences of the potential time-dependency schemes in the flashing ratchet potential, for which the differences are more distinct.  In Fig.~\ref{fig:schemes} we have plotted $v$ and Pe of $N=5$ and $9$ (similar behavior is observed for other values of $N$) RD and FM polymers as a function of $T$ for all three time-dependency schemes.

Some clear differences between the schemes can be seen.  The maxima for the drift and the Peclet numbers are reached for smaller $T$ for type 1 than for types 2 and 3.  The type 2 scheme has the largest $v$ and type 3 the smallest, and the same goes for Pe.  However, this order changes for the inverse drifts, where types 2 and 3 are equally good.  The time-dependency scheme turns out to have an effect on the current inversion phenomena, since the type 3 scheme is able to invert all RD polymers with $N>2$, whereas types 1 and 2 only those with $N>5$.  Despite this, the differences between types 2 and 3 are small (type 2 being slightly ``better'') and we now concentrate only on types 1 and 2.

\subsection{Motor efficiency and stopping force}

\begin{figure}
\includegraphics[width=7.0cm]{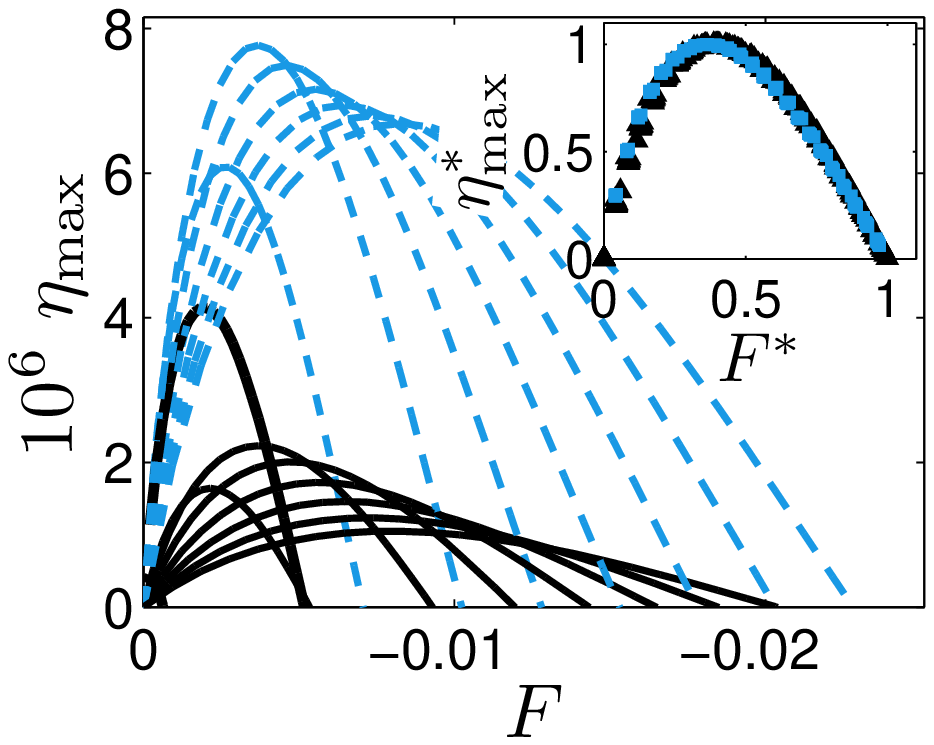}
\includegraphics[width=7.0cm]{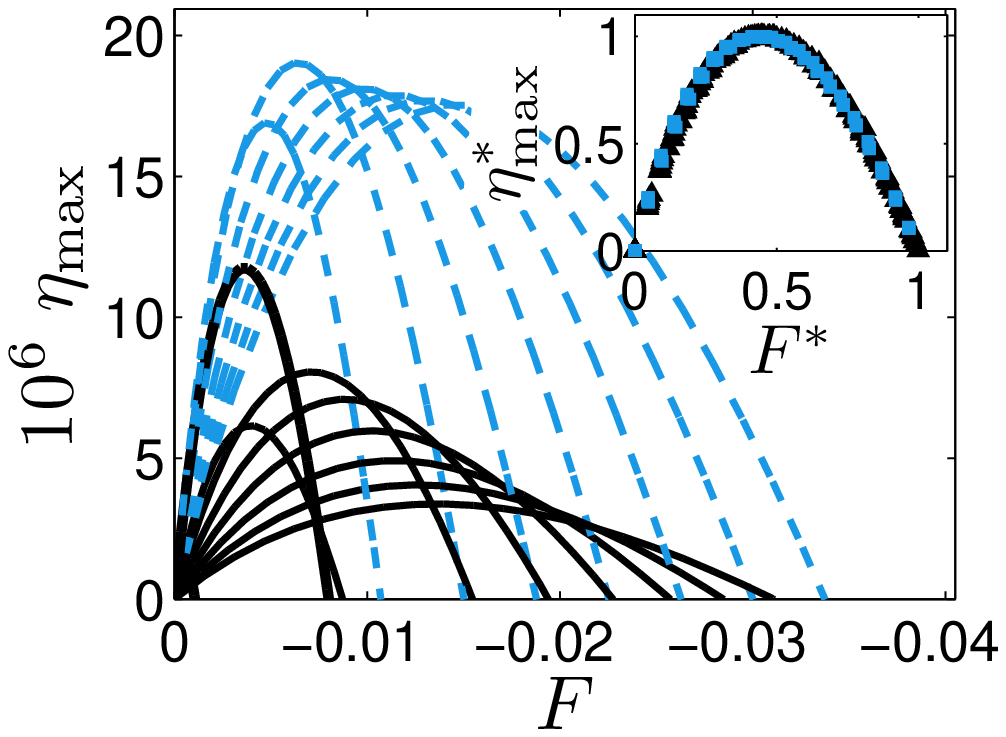}
\caption{(Color online) Maximum efficiency for RD (black solid lines) and FM (blue dashed lines) polymers as a function of the load force $F$ with $N=1...9$ in type 1 (left) and type 2 (right) flashing ratchets.  In each case, the rightmost curve is for $N=9$ and the bold lines (the less interesting special cases $N=1,2$) are shared for both RD and FM polymers. Insets: Rescaled data 
$\eta^*_{\rm max}$ as a function of $F^*$, 
with black triangles for RD polymers and blue squares for FM polymers.}
\label{fig:stopping_ratchet}
\end{figure}

\begin{figure}
\includegraphics[width=7.0cm]{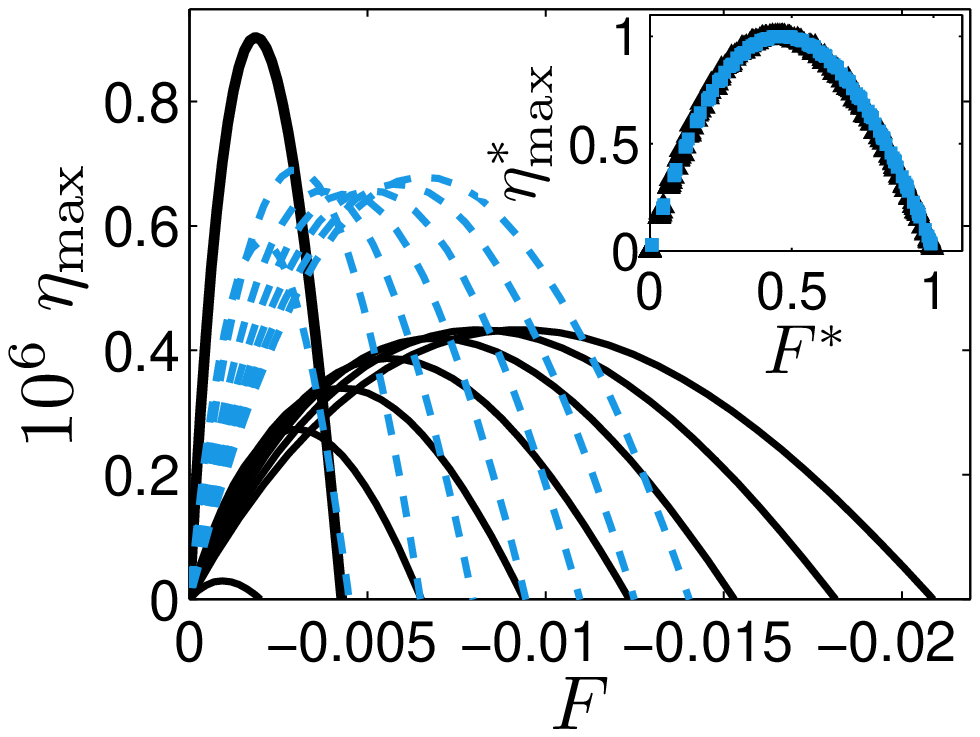}
\includegraphics[width=7.0cm]{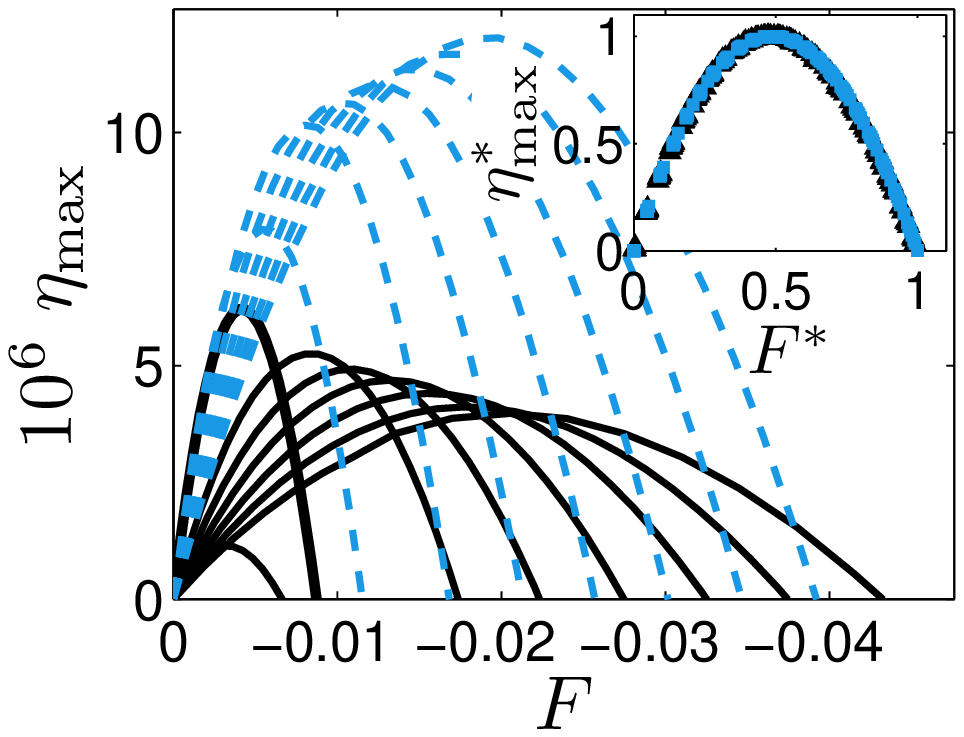}
\caption{(Color online) Maximum efficiency for RD (black solid) and FM (blue dashed) polymers as a function of the load force $F$ with $N=2...9$ in type 1 (left) and type 2 (right) traveling potentials.  In each case, the rightmost curve is for $N=9$ and the bold lines (the less interesting special case $N=2$) are shared for both RD and FM polymers. Insets: Rescaled data 
$\eta^*_{\rm max}$ as a function of $F^*$, with black triangles for RD polymers and blue squares for FM polymers.}
\label{fig:stopping_traveling}
\end{figure}

In Figs.~\ref{fig:stopping_ratchet} and \ref{fig:stopping_traveling} we show the maximum efficiency $\eta_{\rm max} = \max \eta (T)$ of the RD and FM polymers as a function of a load force $F=\sum_{i=1}^{N} E q_i$, where $E$ is the field strength, with flashing ratchets and traveling potentials of the type 1 and 2.  The points where $\eta_{\rm max}(F)=0$ for $F>0$ define the stopping forces $F_{\rm stop}$.  Insets of the figures show the same data scaled with $F^* = F / F_{\rm stop}$ and $\eta_{\rm max}^* = \eta_{\rm max}(F^*) / \max \eta_{\rm max}(F^*)$ for each polymer size, which reveal the shapes of the curves.

We notice that for the FM polymers the efficiency is generally larger and they can maintain their drift in an opposing field better than the RD polymers in the ratchet.  When plotted as a function of $E$, there is a constant stopping field for all $N>3$ FM polymers in both potentials with values around $-0.0026$/$-0.0016$ for type 1 and $-0.0038$/$-0.0043$ for type 2 ratchet/traveling potentials.  This results from the fact that the reptons of the FM polymer are less correlated than those of the RD polymer and the FM polymer thus behave more independently .  For the ratchet, the type 2 scheme is found to be 2-4 times more efficient and can withstand almost double load force when compared with the type 1.  The stopping force is larger for FM polymers.  For the traveling potential, differences are more drastic, as for the type 2 scheme the stopping force is about two times and the efficiency almost one order of magnitude larger when compared to the type 1 scheme.  Rescaled curves reveal that despite the large differences in scales, shape of the curves are almost identical for all polymer lengths and both types.

The numerical values of the efficiency are very small.  This is a generally known trait especially for flashing ratchet models \cite{Cisneros}, but it also results from the choice of the rates, since the velocity plays dominating role for the efficiency.  By the use of the optimized parameters (e.g., $V_{\rm max},x,\mathbf{q}$), efficiency could be increased by couple orders of magnitude.  Results show that $F_{\rm stop}$ increases as a function of $N$, which is in agreement with some previous work \cite{Downton,Kazunarib}.  The efficiency $\eta_{\rm max}$ however decreases as the polymer gets longer for all other but the type 1 traveling ratchet, which is surprising.

\subsection{Non-uniform charge distributions}

Extensive computations were carried out to find the charge configurations with the largest possible $v$ in forward and backward transport and Pe for various polymers and parameters.  It was found that changes in the drift are so large that one can safely limit to maximizing $v$ alone, since in this case Pe is dominated by the drift.  In the following, some of the optimization results are presented for the 8-repton polymers in the type 1 potentials.  The basic model with an uniform charge distribution ($q_i=1\, \forall\, i$) is also shown for comparison.

\begin{figure}
\includegraphics[width=12.0cm]{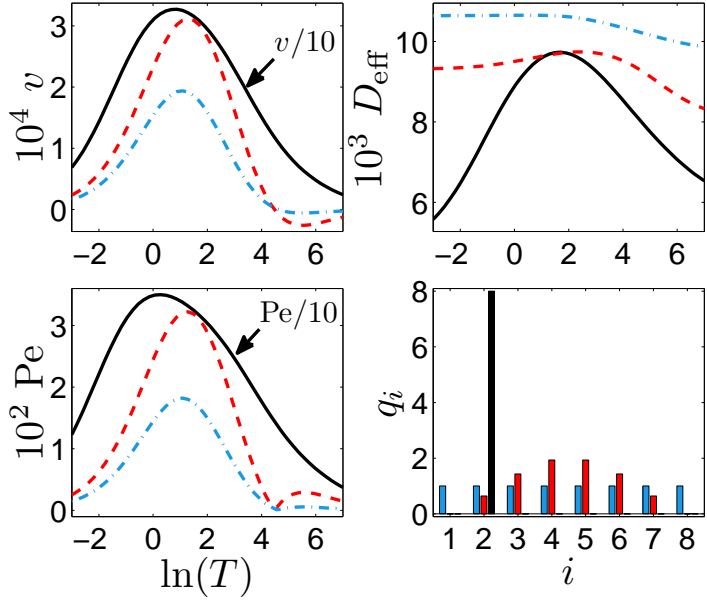}
\caption{(Color online) Drift velocity, diffusion coefficient and Peclet number for the 8-repton RD polymer in the flashing ratchet with uniform (blue dash-dotted lines), negatively optimized (red dashed lines), and positively optimized (black solid lines) charge distributions as a function of the temporal period $T$. The histogram shows the charge distribution along the polymer for each case in the same order. In the leftmost figures $v$ and Pe for the positively optimized case have been scaled by an additional factor 1/10.}
\label{fig:optimal_ratchet}
\end{figure}

In Fig.~\ref{fig:optimal_ratchet}, the properties of the RD polymer in the flashing ratchet are plotted as a function of $T$ with configurations that give maximum drifts for positive (forward) and negative (backward) directions, and the corresponding optimization results are called either positive or negative.  We found that the positive direction is always maximized by putting all charge near either of the heads, but charging the head reptons does not necessarily lead to the largest current. This holds for both RD and FM polymers for all studied polymer lengths up to $N=13$ at least.  In this situation only one repton feels the potential and very large transition rates are generated by the exponential function (see Sec.~II B).  This one repton then forces the whole polymer to advance.

\begin{figure}
\includegraphics[width=10.0cm]{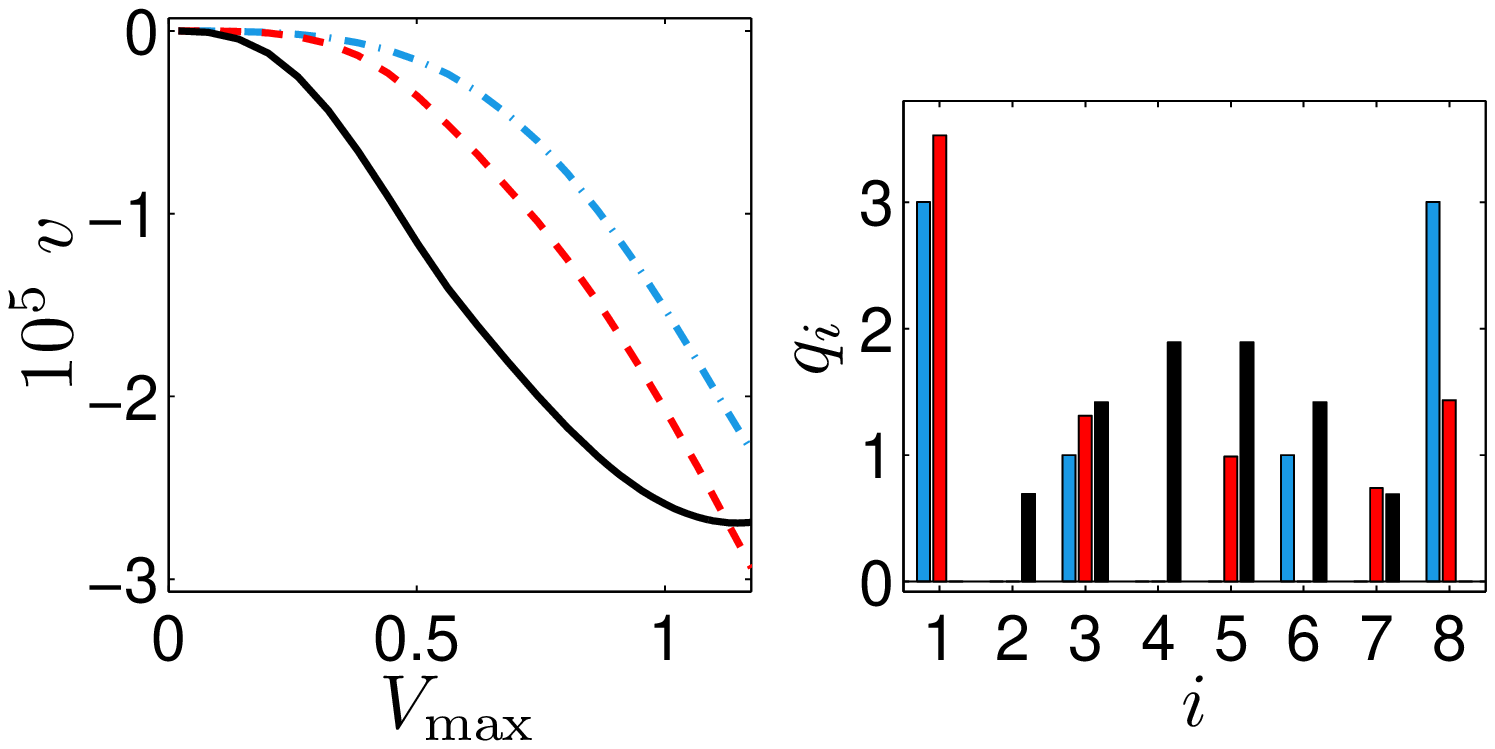}
\caption{(Color online) Left panel: Drifts generated by three locally optimal charge distributions of the 8-repton RD polymer to negative current direction as a function of $V_{\rm max}$.  Right panel: The charge distributions at $V_{\rm max}=1.1734$, with the leftmost bar corresponding to the dash-dotted blue line, the bar in the middle to the dashed red line, and the rightmost bar to the solid black line.}
\label{fig:optimal_phases}
\end{figure}

The optimal charge distributions in the negative direction are more interesting, since the large accumulations of the charge are not seen and the charge is distributed over several reptons.  Symmetric distributions results that neither of the heads are leading and are forced to compete with each other.  This would be very inefficient in constant-field transport.  Repeating the optimization computation several times, additional distributions very close to the first one are found.  Similar local optima are also found in other cases, which complicate the search for the global optimal distribution.  This is demonstrated in Fig.~\ref{fig:optimal_phases}, where we fix $T=\exp(5.5)$ and show the drifts given by the three local optimal distributions as a function $V_{\rm max}$ for the RD polymer in the flashing ratchet.  At $V_{\rm max} \approx 1.126$ the non-symmetric distribution becomes the fastest one.

\begin{figure}
\includegraphics[width=12.0cm]{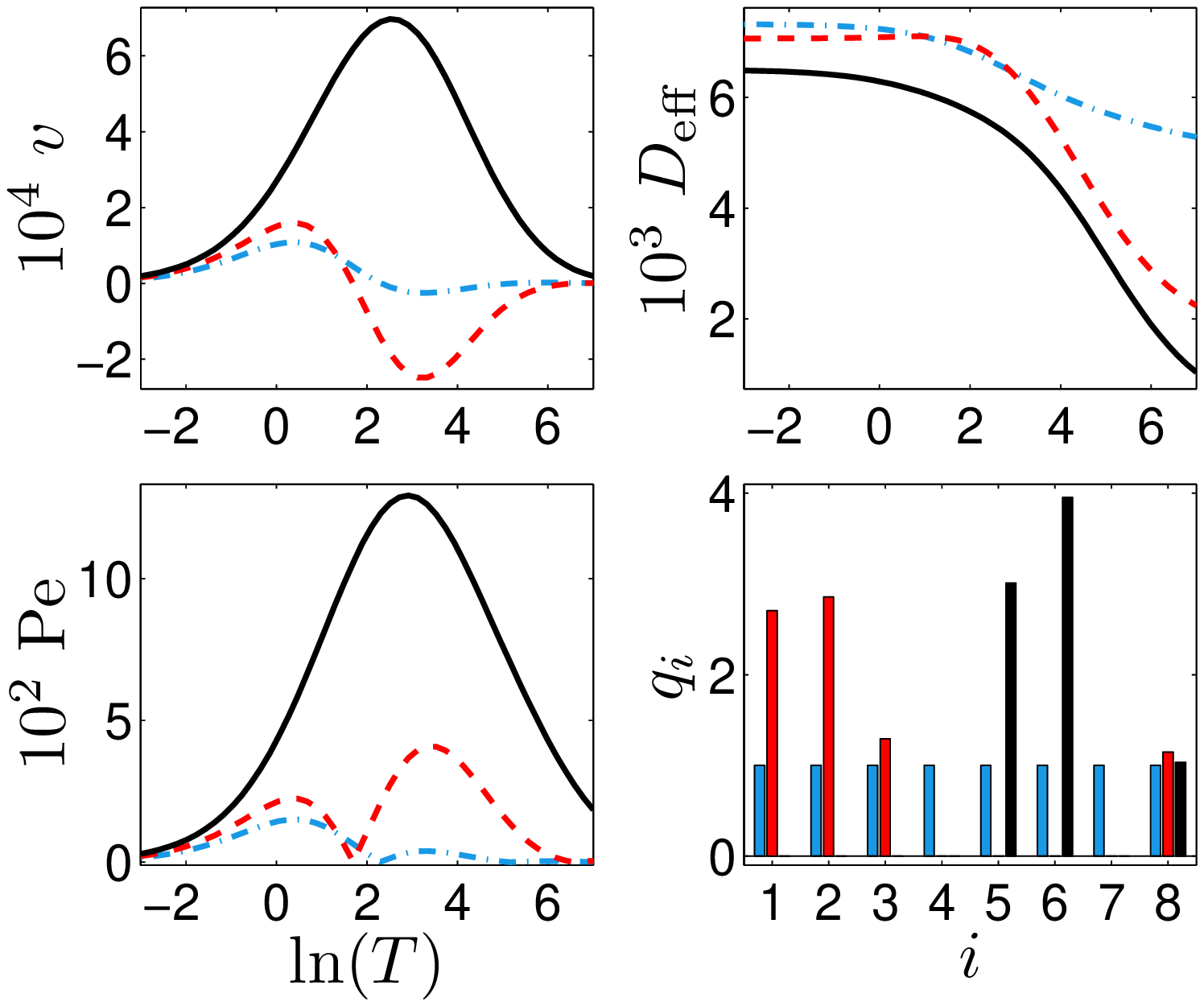}
\caption{(Color online) Drift velocity, diffusion coefficient and Peclet number for the 8-repton RD polymer in the traveling potential with uniform (blue dash-dotted lines), negatively optimized (red dashed lines), and positively optimized (black solid lines) charge distributions.  The histogram shows the charge distribution for each case in the same order.}
\label{fig:optimal_traveling_RD}
\end{figure}

\begin{figure}
\includegraphics[width=12.0cm]{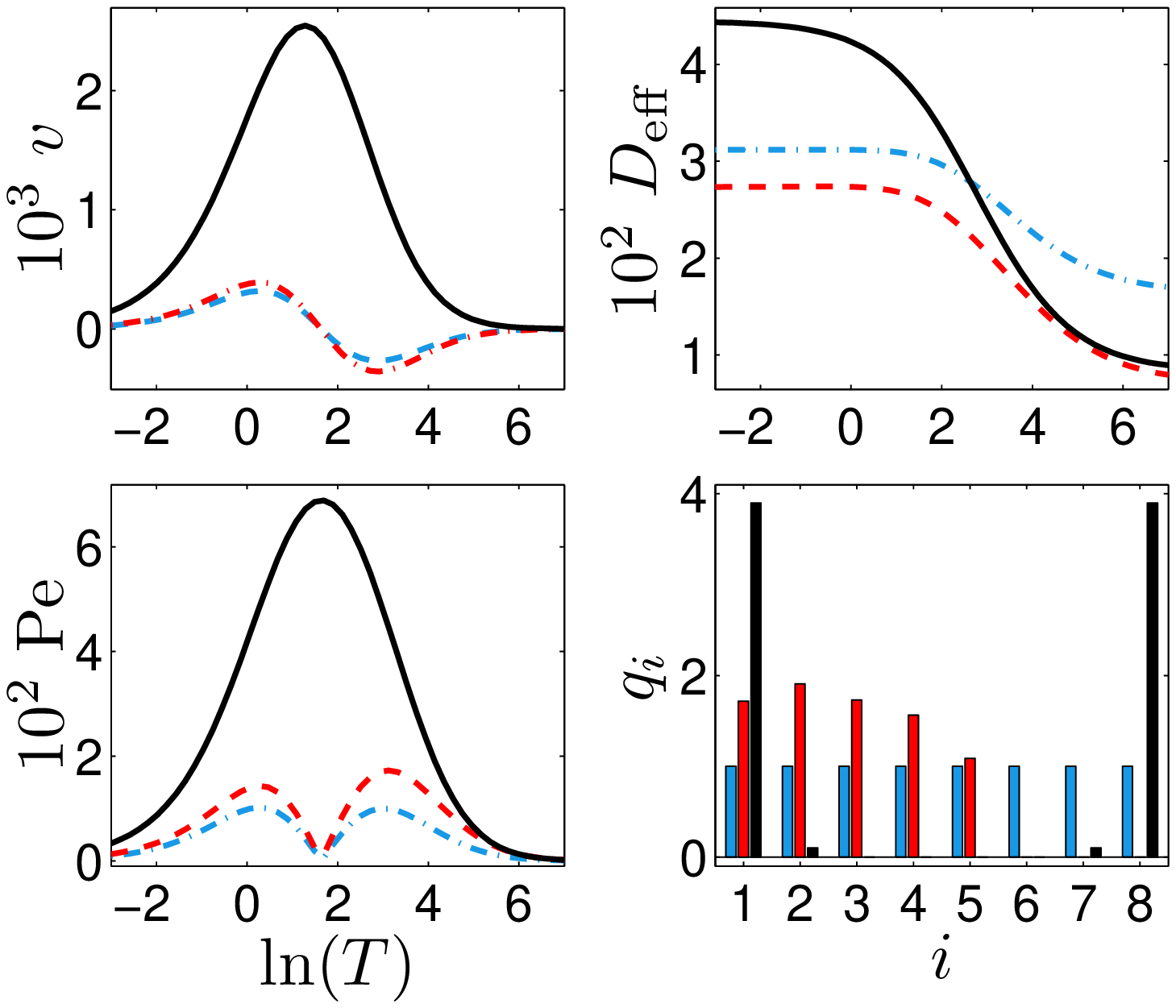}
\caption{(Color online) Drift velocity, diffusion coefficient and Peclet number for the 8-repton FM polymer in the traveling potential with uniform (blue dash-dotted lines), negatively optimized (red dashed lines), and positively optimized (black solid lines) charge distributions.  The histogram shows the charge distribution for each case in the same order.}
\label{fig:optimal_traveling_FM}
\end{figure}

In Figs.~\ref{fig:optimal_traveling_RD} and \ref{fig:optimal_traveling_FM} we show the same analysis for the traveling potential.  For RD polymers, the optimal distributions have no ``clear'' or symmetric structure, only near optimum symmetric distributions are found.  As seen above for the flashing ratchet, the negatively optimized polymers are actually faster in both directions when compared with the uniformly charged polymers.

In conclusion, the charge distribution has a large effect on the polymer transport velocity and coherence on the flashing and traveling potentials.  Since the drifts generated by the ratchet effect are generally very small and difficult to observe, this could be of interest from the point of view of applications.  In the next Section we show that different distributions also lead to different kinds of transport mechanisms.

\section{Results for the internal dynamics of the polymers} % Juha: The internal dynamics of the polymers

\subsection{Time evolution of the observables}

\begin{figure}
\includegraphics[width=12cm]{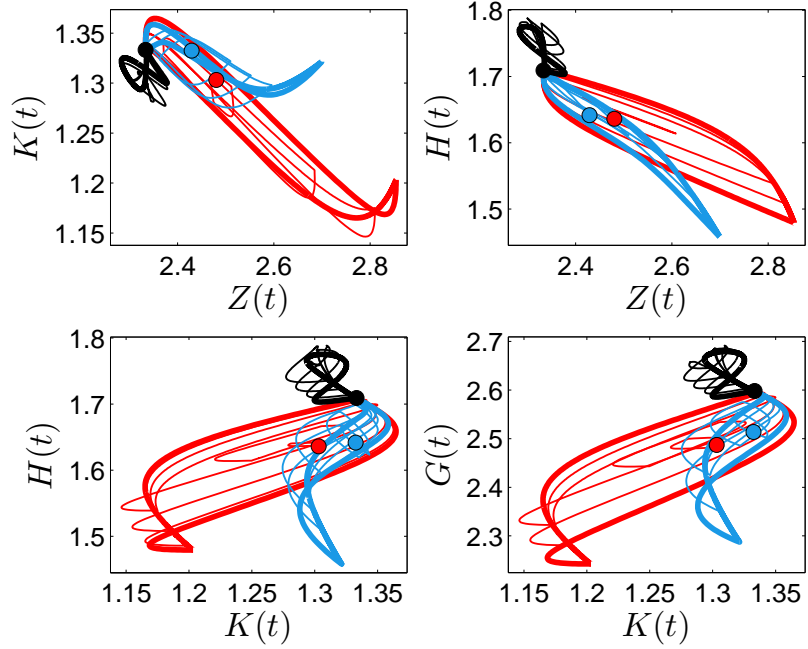}
\caption{(Color online) Time evolution of the 8-repton RD polymer observables in the type 2 flashing ratchet with uniform (red (gray) lines), positively optimized (black lines) and negatively optimized (blue (light gray) lines) charge distributions.  In each case, the big spots correspond to the (mean-field) limit $T \rightarrow 0$, the bold curves show the $T \rightarrow \infty$ limit and the other curves the behavior for a few selected finite values of $T$.}
\label{fig:correlations_ratchet}
\end{figure}

\begin{figure}
\includegraphics[width=12cm]{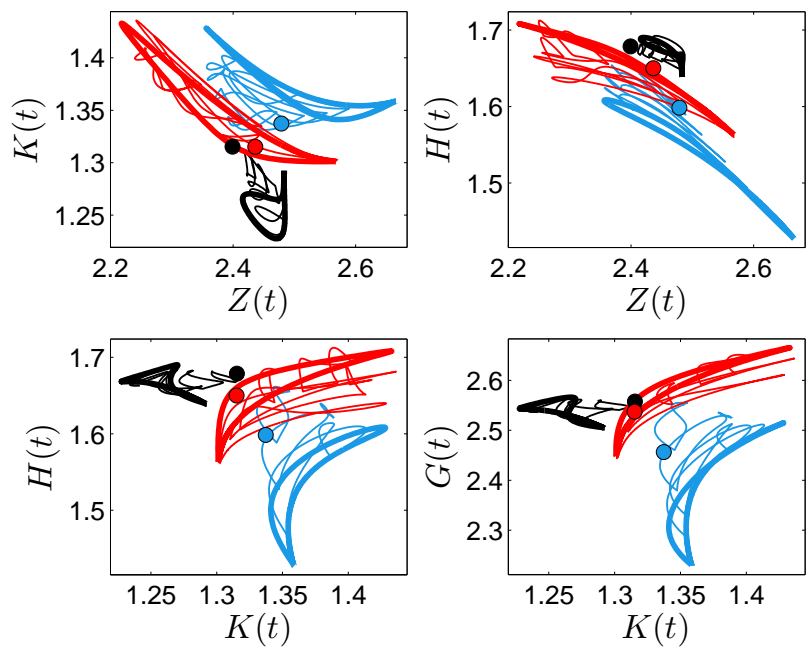}
\caption{(Color online) Time evolution of the 8-repton RD polymer observables in the type 2 traveling potential with uniform (red (gray) lines), positively optimized (black lines) and negatively optimized (blue (light gray) lines) charge distributions.  In each case, the big spots correspond to the (mean-field) limit $T \rightarrow 0$, the bold curves show the $T \rightarrow \infty$ limit and the other curves the behavior for a few selected finite values of $T$.}
\label{fig:correlations_traveling}
\end{figure}

To gain better insight in the internal dynamics of the polymer we now turn to the expected values of the four observables $Z$ (zero-bond), $G$ (total length), $K$ (kinks) and $H$ (head-to-head distance) for the RD polymer.  In Figs.~\ref{fig:correlations_ratchet} and \ref{fig:correlations_traveling} we have plotted the stationary state time-evolution of the observables against each other with the 8-repton RD polymer in the type 2 ratchet and traveling potentials with several values of $T$.  The previously found optimized charge distributions are used.  Note that these distributions are only  \emph{approximately} optimal for the type 2 potentials, but this approximation is found to be very good.  For small $T$, the observables are near their mean-field values (large spots in the figures), which are independent of $t$.  For very large $T$, the curves ``freeze'' (bold lines) since the stationary states are reached before the potential is switched.

For the ratchet, the maximum positive current (black lines) is a result of small changes in the polymer average shape, which is caused by the fact that only a single near-head repton is charged and the rest of the polymer is in pure random motion.  The maximum negative current (blue (light gray) lines) however is a result of more complex processes, which cause much more variation in the average shape, even more than for a polymer with uniformly distributed charges (red (gray) lines) with all the reptons charged.  There is almost one-to-one correlation between $G-H$ pair (as expected), which results that the phase trajectories for the $G-Z$ and $H-Z$ pairs are almost indistinguishable, and therefore the pairs $G-H$ and $G-Z$ are not presented here.  The connections between other pairs are more involved, especially between $H-K$ and $G-K$.  For them, the current inversion is seen as a deformation of loops between $K-Z$, $H-K$ and $G-K$ pairs for uniform and negatively optimized polymers (no current inversion for positively optimized polymer).

For the traveling potential, the curves are more distinguished from each other and are more complicated.  There are fast deformations in the curves as the time goes on.  There is a clear similarity between Figs.~\ref{fig:correlations_ratchet} and \ref{fig:correlations_traveling}.  Positively optimized polymers have the smallest spread in the observables and negatively optimized the largest.  This is similar behavior as seen for the flashing ratchet, albeit the potential and the charge distributions are very different.  The results show that there is a clear connection between the average polymer drift magnitude and direction, and shape deformations.  Deformations during ratcheting for one's part depend strongly on the charge distributions.

\begin{figure}
\includegraphics[width=10cm]{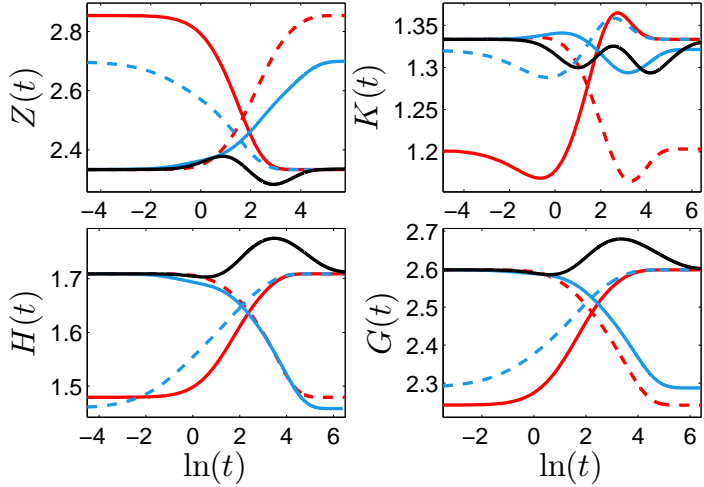}
\caption{(Color online) Relaxation in real time $t$ of the 8-repton RD polymer in the type 2 ratchet potential, with uniform (red (gray) lines), positively (black lines) and negatively (blue (light gray) lines) optimized charge distributions.  Dashed lines (when present) are for the ``on $\rightarrow$ off'' and solid ones for the ``off $\rightarrow$ on'' processes.}
\label{fig:relaxation_ratchet}
\end{figure}

\begin{figure}
\includegraphics[width=10cm]{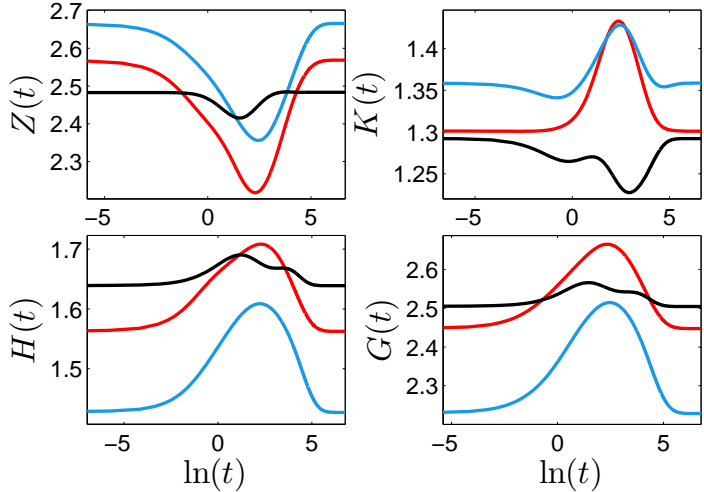}
\caption{(Color online) Relaxation in real time $t$ of the 8-repton RD polymer in the type 2 traveling potential, with uniform (red (gray) lines), positively (black lines) and negatively (blue (light gray) lines) optimized charge distributions.}
\label{fig:relaxation_traveling}
\end{figure}

In Figs.~\ref{fig:relaxation_ratchet} and \ref{fig:relaxation_traveling} we have plotted the relaxation of the observables in the flashing ratchet and the traveling potential for the 8-repton RD polymer with uniform and optimized charge distributions.  The data is the same as shown in Figs.~\ref{fig:correlations_ratchet} and \ref{fig:correlations_traveling} for the large $T$ limit (bold lines).  For the observables, the largest changes are observed in roughly the same time scale, around $\ln (t) \approx 2$.  Stationary values for observables for the positively optimized polymer in the flashing ratchet are independent of the potential state (on or off).   In addition to the kink dynamics, large differences are shown in zero-bond dynamics.  Note that for positively optimized polymer, values remain unchanged during ``on $\rightarrow$ off'' switching and are therefore not shown in the figure.  This is because, in the steady state, the potential has no effect on the conformations of the polymer, which would require more than one charged reptons. For the traveling potential, the time-evolution of the observables is more complex. 

% Chosen T's for the flashing ratchet: T->0 (mean-field), 0.3951, 2.1940, 6.8797, 21.5727, 67.6458, T->inf (asymptotic)
% Chosen T's for the traveling potential: T->0 (mean-field), 0.3951, 2.1940, 6.8797, 21.5727, 119.7869, T->inf (asymptotic)

\subsection{Network analysis}

To further understand the formation of the net drift, we now turn to the network analysis of the steady state currents.  We concentrate on the RD polymer of the type 1 in the flashing ratchet and the traveling potential with uniform and optimized charge distributions.  The temporal periods $T$ are chosen such that they result in the maximum current (4 values of $T$ for both potential types).  The graphs $G^{\rm sign}$ containing the steady-state net currents between the states are then computed.  We have summed over all the potential states ($SL$ degrees of freedom) so that only the internal states of the polymer remain.  After these steps we have eight different graphs with 5832 non-zero directed edges in each of them.

\begin{figure}
\includegraphics[width=9.0cm]{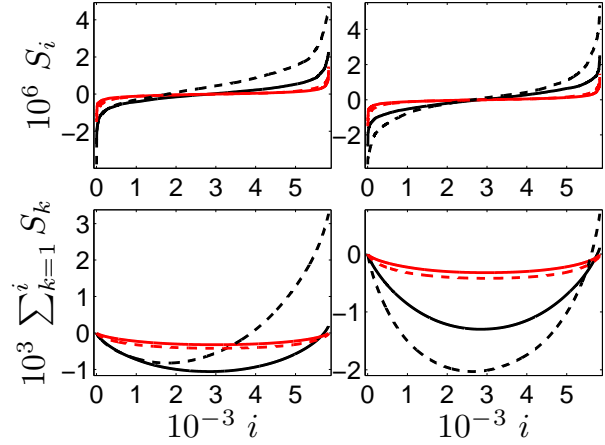}
\caption{(Color online) Upper panels: Ordered elements $S_i$ of the matrix $G^{\rm sign}$ for the $8$-repton RD polymer in the type 1 flashing ratchet (left column) and traveling potential (right column) to positive (black lines) and negative directions (red (gray) lines), with   uniform charge distributions (solid lines) and optimized distributions (dashed lines). Lower panels: The corresponding cumulative sums $\sum_{k=1}^i S_k$.}
\label{fig:ratevertailua}
\end{figure}

Let us first analyze these $G^{\rm sign}$ graphs by defining the arrays $S$ with elements $S_i$ $(i=1...5832)$ that contain all edge weights of the graphs $G^{\rm sign}$ in an increasing order.  In Fig.~\ref{fig:ratevertailua} we show $S_i$ and their cumulative sums.  The total drift $v$ is then recovered as the sum of all $S_i$ (c.f.~Figs.~6 and 8) and a non-zero drift is produced when the $S$ curves are tilted due to the external forces.  Although the system is far from equilibrium, only a slight tilt is observed and there are no single dominating edges.

\begin{figure}
\includegraphics[width=6.0cm]{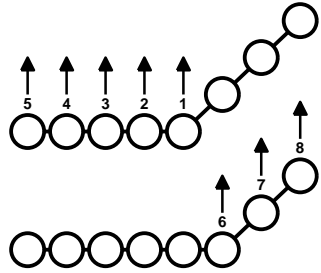}
\caption{Illustration of the positive 4-5-scheme for the 8-repton polymer. The arrows and numbers indicate the direction and the order of the transition for the corresponding reptons.  For clarification, the process is shown here in two parts.}
\label{fig:basic_scheme}
\end{figure}

We now turn to the dominating transport cycles of the polymer motion by analyzing the paths in $G^{\rm sign}$.  This results in cycles with lengths of the order of $10$.  It is found that the common transportation type is such that we call ``$s_1$-$s_2$-scheme'' consisting of cyclically accumulated (lengths $s_1$ and $s_2$ with $|s_1 - s_2|=1$) and elongated parts of the polymer.  Corresponding to the direction of moves, this scheme can be either positive (up) or negative (down).  To illustrate the scheme, we have sketched the positive $4$-$5$ scheme in Fig.~\ref{fig:basic_scheme}.  The numbered arrows indicate the order and direction of the corresponding repton moves.  After all marked moves are done, the initial state is recovered and the cycle is repeated.  In the five situations out of eight studied here, the dominating cycle is the $s_1$-$s_2$-scheme.

\begin{figure}
\includegraphics[width=6.0cm]{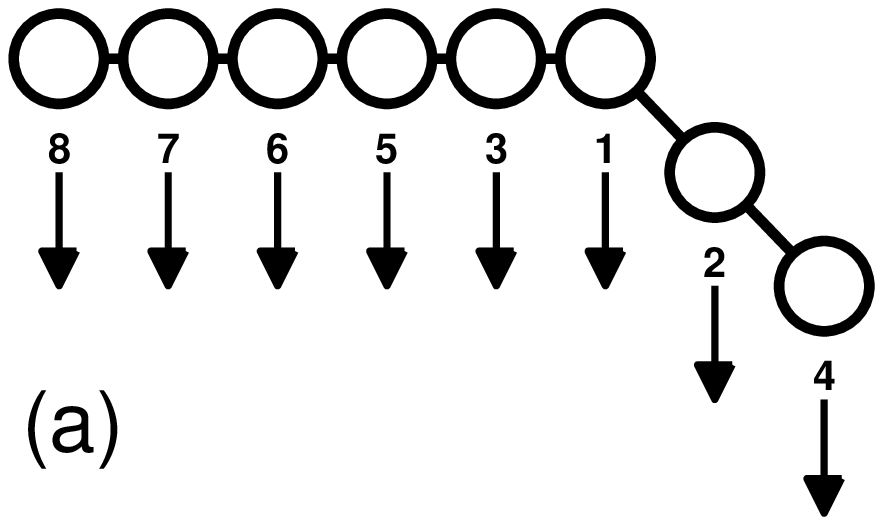}
\includegraphics[width=6.0cm]{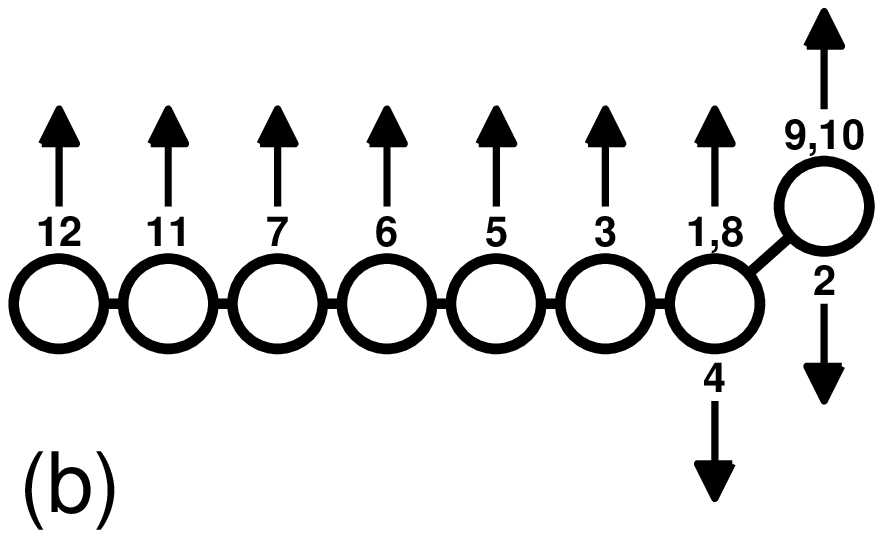}
\includegraphics[width=6.0cm]{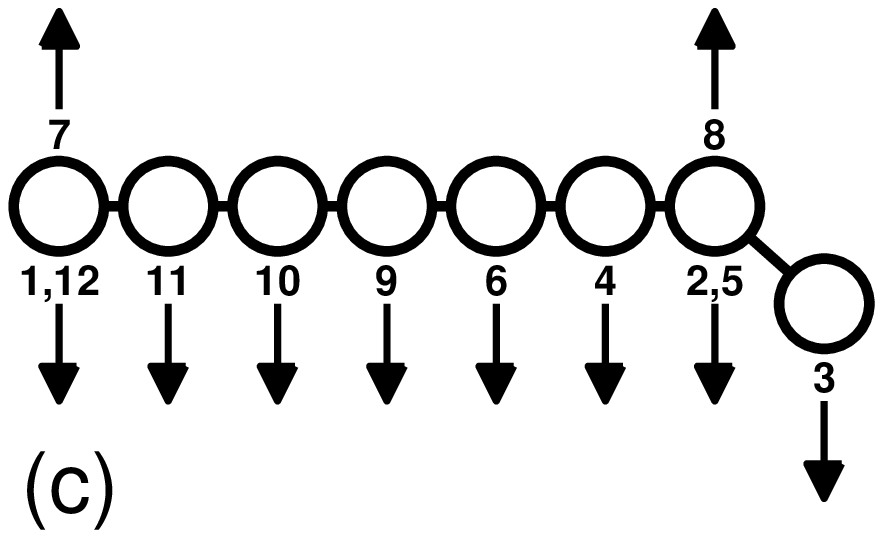}
\caption{Dominating cycles for the backward drift in the traveling potential with the uniform charge distribution (a), the forward drift in the traveling potential with the uniform charge distribution (b), and the backward drift in the ratchet with the optimal charge distribution (c).}
\label{fig:syklit}
\end{figure}

In Fig.~\ref{fig:syklit} we show the remaining three situations that are not of the type above.  Note that for negative transport in the ratchet with the uniform charge distribution, the mechanism is almost the negative 4-5-scheme.

In Table \ref{tbl:network} we have collected the core results of this Section.  For comparison, there is a ratio of the average cycle drift $v_c$, divided by the average drift by the all transitions $v_{\rm all}=v/5832$ in the last column.  This ratio is significantly larger for uniformly charged polymers, indicating that the optimization process increases the drifts for large number of paths and makes differences between paths smaller.  It is also somewhat surprising that there is not much difference between the leading mechanisms for forward or backward motion, and for uniformly charged polymer in ratchet it is actually the same.  One can therefore conclude that the current inversion for the RD model is not caused by some abrupt 'phase transition', but gradual changes in the probability distribution along internal states.

\begin{table}
\centering
\begin{tabular}{lcc}
\toprule
Case & Cycle & $v_{c}$/$v_{\rm all}$ \\ 
\hline
\multicolumn{2}{l}{Ratchet potential} \\
\hline
unif. F & pos. 4-5 & 46,4\\
unif. B & neg. 4-5 & 554\\
optim. F & pos. 2-3 & 6,5\\
optim. B & Fig. \ref{fig:syklit}(c) & 164\\
\hline
\multicolumn{2}{l}{Traveling potential} \\ 
\hline
unif. F & Fig. \ref{fig:syklit}(b) & 60,0\\
unif. B & Fig. \ref{fig:syklit}(a) & 371\\
optim. F & pos. 2-3 & 29,4\\
optim. B & neg. 1-2 & 91,3\\
\end{tabular}
\caption{Dominating cycle types for polymers in ratchet and traveling potentials for forward (F) and backward (B) transport, with uniform (unif.) and optimized (optim.) charge distributions.}
\label{tbl:network}
\end{table}

We carried out a similar analysis also for the full system without summing over $S$ and $L$, in which case cycles have up to 30 states and there are some modifications to the pure $s_1-s_2$ schemes.  However, these cycles are too lengthy to be reported here.  It was found that sometimes summing over the potential states is necessary to find a non-stationary cycle and sometimes the summing leads to stationary cycle.

\section{Discussion}

We have analyzed the properties of Rubinstein-Duke polymers with some modifications, including tube breaking and non-uniform charge distributions, in time-dependent potentials.  The aim of this work was to further study the properties of complex molecules in out-of-equilibrium conditions and especially the ratchet effect.

In the first part of the study, we extended the previous work reported in Ref.~\cite{Kauttonen} by considering deterministic ratcheting mechanisms, the energetic efficiency and optimized charge distributions of the polymers.  It was found that the deterministically flashing potential is superior when compared to a smoothly varying and stochastic potential for velocity, coherence and efficiency.  However, despite ``scaling differences'' in drift and diffusion, the time-dependency scheme seems to have a minor effect on the qualitative results.  By using the stochastic scheme, we computed the optimal charge distributions to maximize the steady-state velocity in flashing ratchets and traveling potentials.  The differences between these and the uniformly charged polymers were found to be drastic.  Changing the charge distribution also changes the mechanism of how the polymer reshapes itself with respect to the potential.

In the second part, the current inversion phenomenon was investigated in detail by using the optimal charge distributions.  The expected values of certain macroscopic observables (e.g., length and zero-bond count) were computed and large differences between differently charged polymers were found.  To find how the polymer actually moves in the non-equilibrium steady state, we proposed a simple graph analysis method to find most probable series of state transitions (=path) based on the probability currents.  For a periodic system such a path is found as a solution of the optimal cycle ratio problem.  This method is suitable in situations where a huge network is generated by some automated fashion or measurements and cannot be analyzed ``manually'' (e.g., Kinesin network in Ref.~\cite{Lipowsky}).  This method was then used to identify the dominating processes of the polymer transport and was found to be very useful to piece together polymer motion.  However, the general usefulness of this analysis depends on the model and it would be of interest to test it for other complex out-of-equilibrium systems and also with non-periodic boundary conditions.

\acknowledgments

This work was supported by the Magnus Ehrnrooth Foundation and the Finnish Academy of Science and Letters. 
We thank Dr.~Otto Pulkkinen for useful discussions.

\appendix

\section{Details of the operators}

In this Appendix the polymer state operators are explained in more detail with some practical aspects of constructing them.

\subsection{Definition of Eq.~(1)}

The explicit definitions of the operators in Eq.~(1) are as in Ref.~\cite{Kauttonen}, but due to the arbitrary charge distributions there is an additional charge dependency in the functions $L$ and $R$:
\begin{multline*}
A_{l}(q) = \{ R(q,l)+L(l) \} \tilde{n}_{\varnothing,1,l} - R(q,l) \tilde{a}_{1,l}^{\dagger}
- L(q,l) \tilde{b}_{1,l}^{\dagger} \\
+ L(q,l) \tilde{n}_{A,1,l} - L(q,l) \tilde{a}_{1,l} + R(q,l) \tilde{n}_{B,1,l} - R(q,l) \tilde{b}_{1,l}
\end{multline*}
\begin{multline*}
B_{y,l}(q) = \{ R(q,l+f(N-1,y))+ L(q,l+f(N-1,y)) \} n_{\varnothing,N-1,y,l} \\ 
- R(q,l+f(N-1,y)) a_{N-1,y,l}^{\dagger} - L(q,l+f(N-1,y)) b_{N-1,y,l}^{\dagger} \\
+ L(q,l+f(N-1,y)) n_{A,N-1,y,l} - L(q,l+f(N-1,y)) a_{N-1,y,l} \\
+ R(q,l+f(N-1,y)) n_{B,N-1,y,l} - R(q,l+f(N-1,y)) b_{N-1,y,l}
\end{multline*} 
\begin{multline*}
M_{i,y,l}(q) = R(q,l+f(i,y))(n_{A,i,y,l}n_{\varnothing,i+1,y,l} + n_{\varnothing,i,y,l}n_{B,i+1,y,l} - a_{i,y,l} a_{i+1,y,l}^{\dagger}-b_{i,y,l}^{\dagger} %%@
b_{i+1,y,l}) \\ 
+ L(q,l+f(i,y))(n_{\varnothing,i,y,l}n_{A,i+1,y,l}+n_{B,i,y,l}n_{\varnothing,i+1,y,l} - a_{i,y,l}^{\dagger} a_{i+1,y,l}-b_{i,y,l} b_{i+1,y,l}^{\dagger}) %%@
\\
+ \Omega R(q,l+f(i,y))(n_{A,i,y,l}n_{B,i+1,y,l} + n_{\varnothing,i,y,l} n_{\varnothing,i+1,y,l} - a_{i,y,l} b_{i+1,y,l} - b_{i,y,l}^{\dagger} %%@
a_{i+1,y,l}^{\dagger}) \\
+ \Omega L(q,l+f(i,y))(n_{B,i,y,l}n_{A,i+1,y,l} + n_{\varnothing,i,y,l} n_{\varnothing,i+1,y,l} - b_{i,y,l} a_{i+1,y,l} - a_{i,y,l}^{\dagger} %%@
b_{i+1,y,l}^{\dagger}) , 
\end{multline*}
where $\Omega = 0$ for RD polymers and $\Omega = 1$ for FM polymers, and
\begin{align*}
\tilde{a}_{1,l} &= c_{l}^{+} a_{1} \: \quad \tilde{a}_{1,l}^{\dagger} = c_{l}^{-} a_{1}^{\dagger} \\
\tilde{b}_{1,l} &= c_{l}^{-} b_{1} \: \quad \tilde{b}_{1,l}^{\dagger} = c_{l}^{+} b_{1}^{\dagger} \\
\tilde{n}_{z,1,l} &= n_{l} n_{z,1} \\
x_{i,y,l} &= n_l \left( \prod_{j=1}^{i-1} n_{g(y,j),j} \right) x_i \\ 
n_{z,i,y,l} &= n_l \left( \prod_{j=1}^{i-1} n_{g(y,j),j} \right) n_{z,i}
\end{align*}
with $x \in  \{ a,b,a^{\dagger},b^{\dagger} \}$, $z \in  \{ A,\varnothing,B \}$.  The function $g(y,i) \in \{ A,\varnothing,B \}$ gives the state of %%@
the $i$th bond in the configuration $y$, and the function $f$ 
\begin{gather*}
f(i,y) =  \sum_{j=1}^{i} \bra{\Psi_{y}} n_{A,i}-n_{B,i} \ket{\Psi_{y}}
,\quad 1 \leq i \leq N-1
\end{gather*}
gives the position of the repton $i+1$ in marker-centered coordinates.  The detailed forms of the functions $g$ and $f$ depend on %%@
the selection of the state basis.  With these functions, the formal definitions for the macrostate observables, i.e., zero-bonds, kinks, head-to-head length and the total length, of the $N$ repton polymer are
\begin{align*}
y \in F_{n_z}^Z: \ & \# \left\{ 1 \leq i<N ;g(y,i) = \varnothing \right\}= n_z, \\
y \in F_{n_k}^K: \ & \# \left\{ 1 \leq i<N-1 ; g(y,i) = \mathrm{A/B} \wedge g(y,i+1) = \mathrm{B/A} \right\}= n_k, \\
y \in F_{n_h}^H: \ & | f(N-1,y) | = n_h, \\
y \in F_{n_g}^G: \ & \max_{k,l} \left[ f(k,y) - f(l,y) \right] = n_{g}, \ k,l=1,2,...,N-1
\end{align*}
where $n_{z},n_{h},n_{g}=0,1,...,N-1$ and $n_{k}=0,1,...,N-2$.  One can verify that $\# F_i^G \geq \# F_i^H$ holds for all $i$.  By using above sets $F$ and equation \eqref{eq:general_op}, measure operands can be constructed and expected values computed.  The practical procedure to form all the required operators, especially the previous observables, is explained below.

\subsection{Operator construction}

Since the stochastic generator and measurement operands used in this work are slightly more complex than in the previous works regarding the RD model, we show in some details how the idea of the recursive operator construction work in the current case.  Whereas small operands can always be build directly, recursive construction is practically a must for large systems and nowadays widely used in DMRG computations \cite{Schollwock,Carlon}.  For simplicity, we concentrate only on (discrete) state measure operators, which in the natural basis are diagonal matrices.

Let $\left\{ O_1^i,...,O_{y_i}^i \right\} $ be a set of macrostate operators for the system with $i$ sites, which includes all the necessary operators that are required when adding a new site.  Here \emph{site} is a general term, which for example could mean single particle states for classical systems and spin states for quantum systems.  By using the usual product state formalism, assume that the new sites are added on the right such that $\ket{\text{new state}} = \ket{\text{old state}} \otimes \ket{\text{new site}}$.  The basic algorithm to add new sites (until $N$) goes as follows\\
\begin{enumerate}
\item Build an initial set of operand(s) $O_{y}^1$, where $y=1,...,y_1$.
\item For all $m=2,3,...,N$ and $y=1,...,y_m$, build: \\ $O_{y}^{m} = \displaystyle \sum_{\begin{subarray}{l} (k,j)=K(y_{m-1},y) \end{subarray} } O_k^{m-1} \otimes \widehat{n}_j$ \\
\item Build the full operand $O^N = \sum_{y=1}^{y_N} \omega_y O_{y}^{N},$
\end{enumerate}
where $y_m$ is the total number of operands required for the size $m$ system.  The details of how to construct a new set of state operators for the enlarged system by joining the states of the new site and the old operators are hidden in the function $K(y_{m-1},y)$.  The complexity of this function and the number of required operators $y_m$ depends on the type of the operand.  Practically it is the $y_m$ that determines the computational effort needed to build large operators, since $K$ is more or less just keeping book of how to join operators.

We now concentrate on the RD-type model for which \emph{sites} mean bond states between the reptons.  The polymer state operators have the following values for $y_m$, given with brief explanations
\begin{itemize}
\item Zero-bonds: $y_{m}=m+1$ (number of zero-bonds)
\item Kinks: $y_m=\max \left\{ 3,3\left( m-1 \right) \right\}$ (number of kinks and state of the rightmost bond)
\item Head-to-head length: $y_{m}=2m+1$ (signed distance between the heads)
\item Total length: $y_m=(1-I) (3I-2m-7) \geq \left( 1 + m \right) \left( 3 + m \right) / 3$, where $I=\left\lceil \frac{m+4}{3}\right\rceil$ (see example below)
\end{itemize}
The number of required operands is therefore $\propto m^2$ for total length and $\propto m$ for others.

We now consider a concrete example for a total length operator, which is the most complex operator used in this paper.  When one enlarges the size of this operator with new particles, one must keep track of the maximum distances of the rightmost repton from all the other reptons.  For example, in Fig.~\ref{fig:esimerkkikuva} these distances would be 2 (from repton nr. 4) and 0 (no any reptons below).  We define these as \emph{up} (u) and \emph{down} (d) distances.  Total distance is then $d+u$.

\begin{figure}
\includegraphics[width=9.0cm]{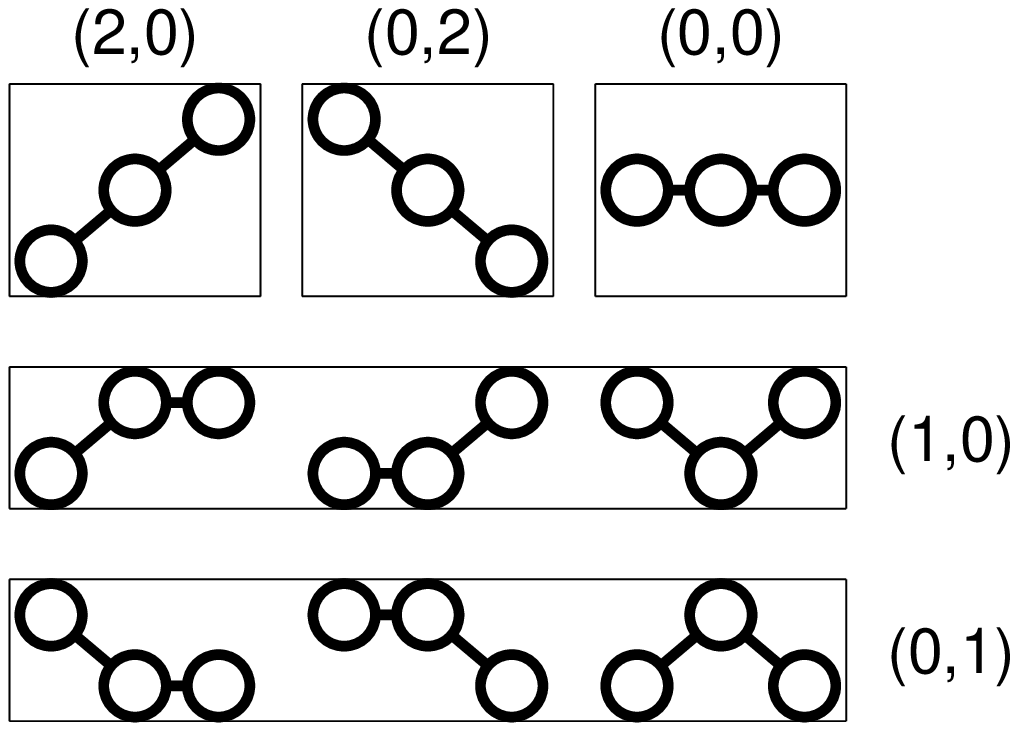}
\includegraphics[width=6.0cm]{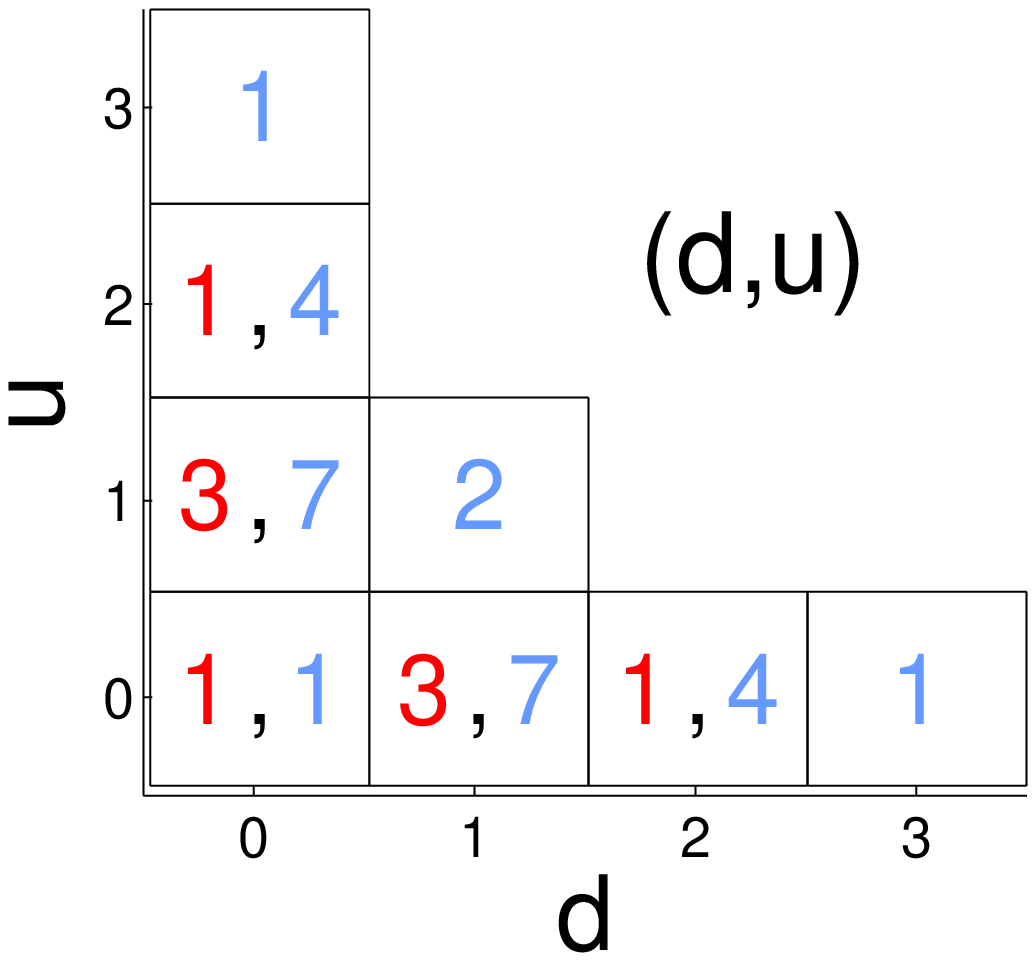}
\caption{(Color online) All 9 configurations (microstates) of the 3-repton polymer with 5 macrostate operators that are formed from them.  The macrostates are indexed by the corresponding $(d,u)$.  As a new particle is added, there are new $(d,u)$ combinations available and the number of macrostates is increased by three.  The relation between the microstates and macrostates is illustrated in the $(d,u)$-plane, where the red (gray) and blue (light gray) numbers indicate the number of microstates for $3$ and $4$ repton polymers.}
\label{fig:appendix_esimerkki}
\end{figure}

In Fig.~\ref{fig:appendix_esimerkki} we show all 9 microstates of the $3$-repton polymer.  Since there are five combinations for $u$ and $d$ distances, the macrostate operators $O_{(2,0)}^2,O_{(0,2)}^2,...,O_{(0,1)}^2$ are formed with each of them including one or more microstates.  This is shown in the figure with red numbers in $(d,u)$-plane.  When a new repton is added, function $K$ is used to combine old macrostate operators with state operators of the new site ($\left\{ n_A,n_{\varnothing},n_B \right\}$) and hence the number of macrostate operators is increased by three.  Examples of the required operations includes $O_{(2,0)}^3 = O_{(1,0)}^2 \otimes n_A$ and $O_{(1,1)}^3 = O_{(2,0)}^2 \otimes n_B$.  After addition, there are 27 microstates in eight macrostate operators (blue numbers in the figure).  Note that in the actual computations only the information about the $d$ and $u$ values is needed.  Here the tracking of the microstates was done for illustration purposes only.  As more reptons are added, the ``triangle'' that presents available $(d,u)$ states gets larger.

\end{document}